\documentclass[useAMS,usenatbib]{mn2e}
\usepackage{txfonts}
\usepackage{natbib}
\usepackage{aas_macros}
\usepackage[all]{xy}

\usepackage{graphicx}

\usepackage{hyperref}
\usepackage[usenames,dvipsnames]{color}
\definecolor{darkblue}{rgb}{0,0,.5}
\hypersetup{bookmarksopen,
			pdfstartview={FitH},
			colorlinks=true,
			breaklinks=true,
			linkcolor=darkblue,
			menucolor=darkblue,
			urlcolor=Red,
			citecolor=Green,
			linkcolor=blue}
\usepackage[all]{hypcap}

\def\del#1{{}}

\newcommand{\dd}{\mathrm{d}}
\newcommand{\eqref}[1]{(\ref{#1})}
\newcommand{\x}{\bmath x}
\newcommand{\kvec}{\bmath k}
\newcommand{\n}{\bmath{\hat{n}}}
\newcommand{\im}{\mathrm{i}}

\title[Nonlinear iSW-effect]{Contributions to the nonlinear integrated Sachs-Wolfe effect: Birkinshaw-Gull effect and gravitational self-energy density}
\author[Philipp M. Merkel and Bj{\"o}rn Malte Sch\"afer]
{Philipp M. Merkel$^1$\thanks{e-mail: philipp.merkel@urz.uni-heidelberg.de} and Bj{\"o}rn Malte Sch\"afer$^2$\\
${}^1$Institut f{\"u}r Theoretische Astrophysik, Zentrum f{\"u}r Astronomie, Universit{\"a}t Heidelberg, Albert-Ueberle-Stra{\ss}e 2, 69120 Heidelberg, Germany\\
${}^2$Astronomisches Recheninstitut, Zentrum f{\"u}r Astronomie, Universit{\"a}t Heidelberg, M{\"o}nchhofstra{\ss}e 12, 69120 Heidelberg, Germany}

\begin{document}
\pagerange{\pageref{firstpage}--\pageref{lastpage}}
\pubyear{2010}
\maketitle
\label{firstpage}

\begin{abstract}
In this paper, we recompute contributions to the spectrum of the nonlinear integrated Sachs-Wolfe (iSW)/Rees-Sciama effect in a dark energy cosmology. Focusing on the moderate nonlinear regime, all dynamical fields involved are derived from the density contrast in Eulerian perturbation theory. Shape and amplitude of the resulting angular power spectrum are similar to that derived in previous work. With our purely analytical approach we identify two distinct contributions to the signal of the nonlinear iSW-effect: the change of the gravitational self-energy density of the large scale structure with (conformal) time and gravitational lenses moving with the large scale matter stream. In the latter we recover the Birkinshaw-Gull effect. As the nonlinear iSW-effect itself is inherently hard to detect, observational discrimination between its individual contributions is almost excluded. 
Our analysis, however, yields valuable insights into the theory of the nonlinear iSW-effect as a post-Newtonian relativistic effect on propagating photons.
\end{abstract}


\section{Introduction}
\label{sec_introduction}

The tiny fluctuations in the temperature of the cosmic microwave background (CMB), its primary anisotropies, reflect the state of the Universe approximately 400,000 years after the big bang. Photons emanating from the last scattering surface, however, interact on their way to today's observer with the intervening large scale structure (LSS), thereby altering this picture of the early Universe. One of these secondary anisotropies is the integrated Sachs-Wolfe (iSW) effect \citep{1967ApJ...147...73S}. The interaction of CMB photons with time-evolving gravitational potentials in the LSS causes additional fluctuations in the CMB temperature. While being blue-shifted when entering a potential well the photons also experience a redshift on their way out of the potential. A net change in temperature results when blueshift and redshift do not compensate.

Being a secondary effect, the signal of the iSW-effect is quite small in comparison to the CMB temperature itself. Its importance, however, arises from its dedicated sensitivity to cosmic fluids with non-vanishing equation of state \citep{1996PhRvL..76..575C} promoting it as a valuable tool in the investigation of dark energy and non-standard cosmologies \citep{2004PhRvD..69l4015L,2006PhRvD..73l3504Z}, as well as the Universe's spatial curvature \citep{1994ApJ...432....7K}. 

Exploiting its strong correlation with the galaxy distribution the iSW-effect has been detected with high significance by various groups \citep{2004Natur.427...45B, 2006MNRAS.365..891V, 2007MNRAS.376.1211M, 2008PhRvD..77l3520G}. 
The constraints they derive for the matter density $\Omega_{\mathrm m}$ and the dark energy equation of state $w$ give support to the $\Lambda$CDM model, which describes a spatially flat universe mainly constituted by cold dark matter (CDM) and cosmological constant $\Lambda$.

The nonlinear iSW- or Rees-Sciama (RS-) effect \citep{1968Natur.217..511R,1996ApJ...460..549S,2002PhRvD..65h3518C,2006MNRAS.369..425S} refers to any temperature anisotropy caused by nonlinearly evolving gravitational potentials. In contrast to the linear iSW-effect it is a small scale phenomenon dominating the total iSW signal on multipoles $\ell \gtrsim 100$ but is nevertheless hard to detect because of the primary CMB \citep{2002PhRvD..65h3518C,2011MNRAS.416.1302S}. 

Resulting from nonlinear structure formation the statistics of the RS-effect is necessarily non-Gaussian. 
The non-Gaussianities induced by the RS-effect, however, are small. \citet{1999PhRvD..59j3001S} showed that the CMB bispectrum due to the RS-effect is undetectable. Also mixed polyspectra of the nonlinear iSW-effect temperature perturbation and the tracer galaxy density field, as recently investigated by \citet{2012arXiv1210.7513J} in continuation of the work of \citet{2008MNRAS.388.1394S}, do not reach sufficiently high signal-to-noise ratios.
Therefore, stacking methods \citep{2008ApJ...683L..99G} as an alternative to a statistical detection of the RS-effect are of great interest.

\citet{1996ApJ...460..549S} and \citet{2002PhRvD..65h3518C} provided two different perturbative descriptions of the nonlinear iSW-effect yielding comparable results.
In this work we take up the ansatz of \citet{2002PhRvD..65h3518C}. While his results make extensive use of 
findings from the halo model, we are more interested in an analytical treatment. Therefore, we concentrate our analysis on the weakly nonlinear regime where the statistics of the underlying fields are not too far from being Gaussian and second order perturbation theory is well applicable. We compute the nonlinear iSW-effect using solely the density field and its derivatives. Although the applicability of this ansatz is naturally limited, it allows for deeper insights into the underlying physical processes. 
Our approach is well suited for identifying the distinct contributions to the nonlinear iSW-effect as well as 
elucidating their physical origin.

This article is structured in the following way: We begin with a short compilation of the key formulae describing growth and statistics of the density contrast in dark energy cosmologies (Section~\ref{sec_cosmology}). In Section~\ref{sec_non_linear_isw} we compute the spectrum of the nonlinear iSW-effect. Furthermore, we derive the resulting two-point statistics and present the corresponding angular power spectrum. We extend our analysis in Section~\ref{sec_contributions}, where we discriminate and interpret the different contributions to the nonlinear iSW-effect. In Section~\ref{sec_physical_interpretation} the physical interpretation is continued and deepened by pointing out analogies with classical field theory as well as the theory of gravitomagnetic potentials. Finally, we summarize our results in Section~\ref{sec_summary}.

Throughout this work we choose a spatially flat $w$CDM cosmology as reference. The initial perturbations in the CDM component are assumed to be adiabatic and Gaussian distributed with variance $\sigma_8=0.8$, where we set the spectral index $n_s$ to unity. The matter content is described by $\Omega_{\mathrm m}=0.25$ and $\Omega_{\mathrm b}=0.04$ while the value of the Hubble constant is set to $H_0=100 \, h \,\mathrm{km} / \mathrm{s} /\mathrm{Mpc}$ with $h=0.72$.

\section{Cosmology}
\label{sec_cosmology}
\subsection{Dark energy cosmologies}
In a spatially flat Friedmann-Lema\^itre-Robertson-Walker universe the time evolution of the scale factor $a$ is governed by the the Hubble function $H(t)=\dd \log a / \dd t$:
\begin{equation}
 \frac{H^2(a)}{H^2_0} = \Omega_{\mathrm m} a^{-3} + (1 - \Omega_{\mathrm m})
 \exp \left( \int_a^1 \dd \log a\;  \left[1+w(a) \right]\right).
\end{equation}
The expansion of the homogeneous background is thus completely described by the matter content $\Omega_{\mathrm m}$ (in units of the critical density) and the equation of state of the dark energy fluid $w(a)$. Following \citet{2001IJMPD..10..213C} the latter can be phenomenologically parametrized as 
\begin{equation}
 w(a)=w_0 + (1-a)w_a,
\end{equation}
covering a huge variety of different dark energy models. Setting $w_0=-1$ and $w_a=0$ one recovers the equation of state of a cosmological constant.

Since photons follow null geodesics, conformal time $\dd \eta = \dd t/a$ and comoving distance $\chi$ can be used interchangeably allowing for a simple relation to the scale factor:
\begin{equation}
 \chi = c\int_a^1\frac{\dd a}{a^2 H(a)}.
\end{equation}
Naturally, (comoving) distances are measured in units of the Hubble distance $\chi_H=c/H_0$ which sets the scale up to which Newtonian gravity is applicable.

\subsection{CDM power spectrum and linear structure growth}
The fluctuations in the cold dark matter (CDM) component are described by the density contrast
\begin{equation}
 \delta(\bmath x) \equiv \frac{\rho_{\mathrm m}(\bmath x) - \bar{\rho}_{\mathrm m}}{\bar{\rho}_{\mathrm m}}
 \qquad
 \mathrm{with}
 \qquad
 \bar\rho_{\mathrm{m}} = \Omega_{\mathrm m}\rho_{\mathrm{crit}}.
\end{equation}
In linear theory $\delta$ is a statistically homogeneous and isotropic Gaussian random field and thus completely characterized by its power spectrum
\begin{equation}
 \left\langle \delta(\bmath k ) \delta^*(\bmath k')\right\rangle = (2\pi)^3 \delta_D(\bmath k - \bmath k')P_{\delta\delta}(k).
\end{equation}
For the power spectrum we choose a power law modulated by an appropriate transfer function, i.e.
\begin{equation}
 P_{\delta\delta}(k) \propto k^{n_s} T^2(k),
\end{equation}
for which we use the fit proposed by \citet{1986ApJ...304...15B}
\begin{eqnarray}
 \nonumber
  T(q) &=& \frac{\log (1+2.34 q)}{2.34 q} \left(1 + 3.89 q + (16.1q)^2 + (5.46q)^3 \right. \\
  	&&  \left.+ (6.71q)^4\right)^{-\frac{1}{4}}.
\end{eqnarray}
In this formula the wavenumber $q=k\Gamma$ is rescaled by the shape parameter  
\begin{equation}
 \Gamma = \Omega_{\mathrm m} h \exp\left[ -\Omega_{\mathrm b}\left(1 + \frac{\sqrt{2h}}{\Omega_{\mathrm m}}   \right) \right]
\end{equation}
in order to account for a non-vanishing baryon density $\Omega_{\mathrm{b}}$ \citep{1995ApJS..100..281S}. Finally, the amplitude of the power spectrum is set by the variance of the density contrast $\sigma^2_R$ smoothed on the scale $R=8\,\mathrm{Mpc}/h$
\begin{equation}
 \sigma^2_R = \int_0^\infty \frac{\dd k}{2\pi^2} k^2 P_{\delta\delta}(k)W^2_R(k) =
 \int_0^\infty \dd\log k\; \Delta^2(k) W^2_R(kR) ,
\end{equation}
where $W_R(y)=3j_1(y)/(y)$ is the Fourier-transform of a spherical top hat function
\citep[$j_\ell(x)$ denoting the $\ell$-th spherical Bessel function of the first kind,][]{1972hmfw.book.....A}.

During linear evolution the time dependence of the density contrast is completely encapsulated in the growth function $D_+(\eta)$ so that $\delta(\kvec, \eta) = D_+(\eta)\delta_0(\kvec)$ (normalized to unity today). It is obtained by solving the growth equation \citep{1997PhRvD..56.4439T,1998ApJ...508..483W,2003MNRAS.346..573L}
\begin{equation}
 \frac{\dd^2}{\dd a^2}D_+(a) + \frac{1}{a}\left( 3 + \frac{\dd \log H}{\dd \log a}\right) \frac{\dd}{\dd a} D_+(a) = \frac{3}{2a^2}\Omega_{\mathrm m}(a) D_+(a).
\end{equation}

\section{Nonlinear integrated Sachs-Wolfe effect}
\label{sec_non_linear_isw}

The relative change in temperature CMB photons experience when they traverse time-evolving gravitational potentials on their way from the last scattering surface to today's observer along the direction $\bmath{\hat{n}}$ is given by the line of sight integral \citep{1967ApJ...147...73S}
\begin{equation}
 \label{eq_isw_temperature_shift}
 \tau(\bmath{\hat{n}}) \equiv  \frac{\Delta T_{\mathrm{iSW}}(\n)}{T_{\mathrm{CMB}}}=\frac{2}{c^2}\int_0^{\chi_H}\dd \chi \frac{\partial}{\partial\eta}\Phi(\n,\eta).
\end{equation}
The gravitational potential is directly related to the density fluctuations via the (comoving) Poisson equation
\begin{equation}
 \label{eq_Poisson_equation}
   \Delta \Phi (\x,\eta) = \frac{3}{2} H_0^2 \Omega_{\mathrm m} \frac{\delta (\x,\eta)}{a}.
\end{equation}
Its time derivative can be most easily accessed in Fourier space where the Poisson equation is readily inverted
\begin{equation}
 \frac{\partial}{\partial \eta} \Phi (\kvec, \eta) = - \frac{3}{2} H_0^2 \Omega_{\mathrm m} \frac{1}{k^2} \frac{\partial}{\partial \eta}\frac{\delta(\kvec,\eta)}{a}.
\end{equation}
In the derivation of the linear iSW-effect one usually inserts $\delta(\kvec,a)=D_+(a)\delta_0(\kvec)$ revealing that during matter domination where $D_+(a)=a$ the iSW-effect vanishes. For our purposes, however, it proves advantageous to carry out the time derivative formally
\begin{equation}
 \frac{\partial}{\partial \eta} \Phi (\kvec, \eta) =  \frac{3}{2} H_0^2 \Omega_{\mathrm m} \frac{1}{k^2} \left( H(\eta)\delta(\kvec,\eta) 
 -\frac{1}{a}\frac{\partial}{\partial \eta}\delta(\kvec,\eta)\right).
\end{equation}
Obviously, the time evolution of the potential is sourced by two different effects: on the one hand the expansion of the homogeneous background, represented by the first term, and on the other hand by the time varying structure growth. As mentioned before, in epochs dominated by a fluid with vanishing equation of state $(w=0)$ these two effects just balance.

Since we are interested in nonlinear contributions to the iSW-effect we express the time derivative of the density fluctuations by means of the continuity equation.
\begin{equation}
 \frac{\partial}{\partial \eta} \delta(\x,\eta) + \mathrm{div}\, \bmath j(\x,\eta) = 0.
\end{equation}
Due to matter conservation, the change of the dark matter density in time needs to be compensated by the divergence of the corresponding flux
\begin{equation}
 \bmath j(\x,\eta) = [1+\delta(\x,\eta)] \bupsilon(\x,\eta).
\end{equation}
In the linear regime $\delta \ll 1$, thus the momentum density $\delta(\x,\eta) \bupsilon(\x,\eta)$ is negligible and one can immediately read off the familiar Fourier representation of the peculiar velocity field
\begin{equation}
 \label{eq_velocity_density_relation}
 \bupsilon(\bmath k, \eta) = \int \dd^3x\; \bupsilon(\x,\eta) \mathrm{e}^{\mathrm{i}\bmath k \cdot \x} =
-\im \frac{\partial D_+(\eta)}{\partial\eta} \frac{\kvec}{k^2} \delta_0(\kvec).
\end{equation}
In Fourier space, the dark matter current density becomes a convolution
\begin{equation}
 \label{eq_flux_in_Fourier_space}
 \bmath j ( \kvec, \eta) = \bupsilon ( \kvec, \eta) + \int \frac{\dd^3 k'}{(2\pi)^3} \bupsilon (\kvec-\kvec', \eta ) \delta(\kvec', \eta),
\end{equation}
so that the time evolution of the gravitational potential finally can be written as a function of density contrast and peculiar velocities
\begin{eqnarray}
	 \nonumber
	 \frac{\partial}{\partial \eta} \Phi (\kvec, \eta) &=& \frac{3}{2} H_0^2 \Omega_{\mathrm m} \frac{1}{k^2}
	 	\bigl[ H(\eta)\delta(\kvec,\eta) - \im \kvec \cdot 
		\bigl( \bupsilon ( \kvec, \eta) 
		\bigr.
		\bigr.\\
		&&
		\biggl.
		\biggl.
		+ \int \frac{\dd^3 k'}{(2\pi)^3} \bupsilon (\kvec-\kvec', \eta) \delta(\kvec',\eta)
		\bigr)
		\bigr].
\end{eqnarray}
Due to the symmetry of the convolution the expression for the current density is not unique. For example, one could have explicitly symmetrized equation~\eqref{eq_flux_in_Fourier_space}. However, for our purpose, aiming at the different contributions to the RS-effect, the form given above is most useful as will become clear below.

\begin{figure}
 \label{fig_growth_function_and_derivatives}
 \centering
 \resizebox{\hsize}{!}{\includegraphics[]{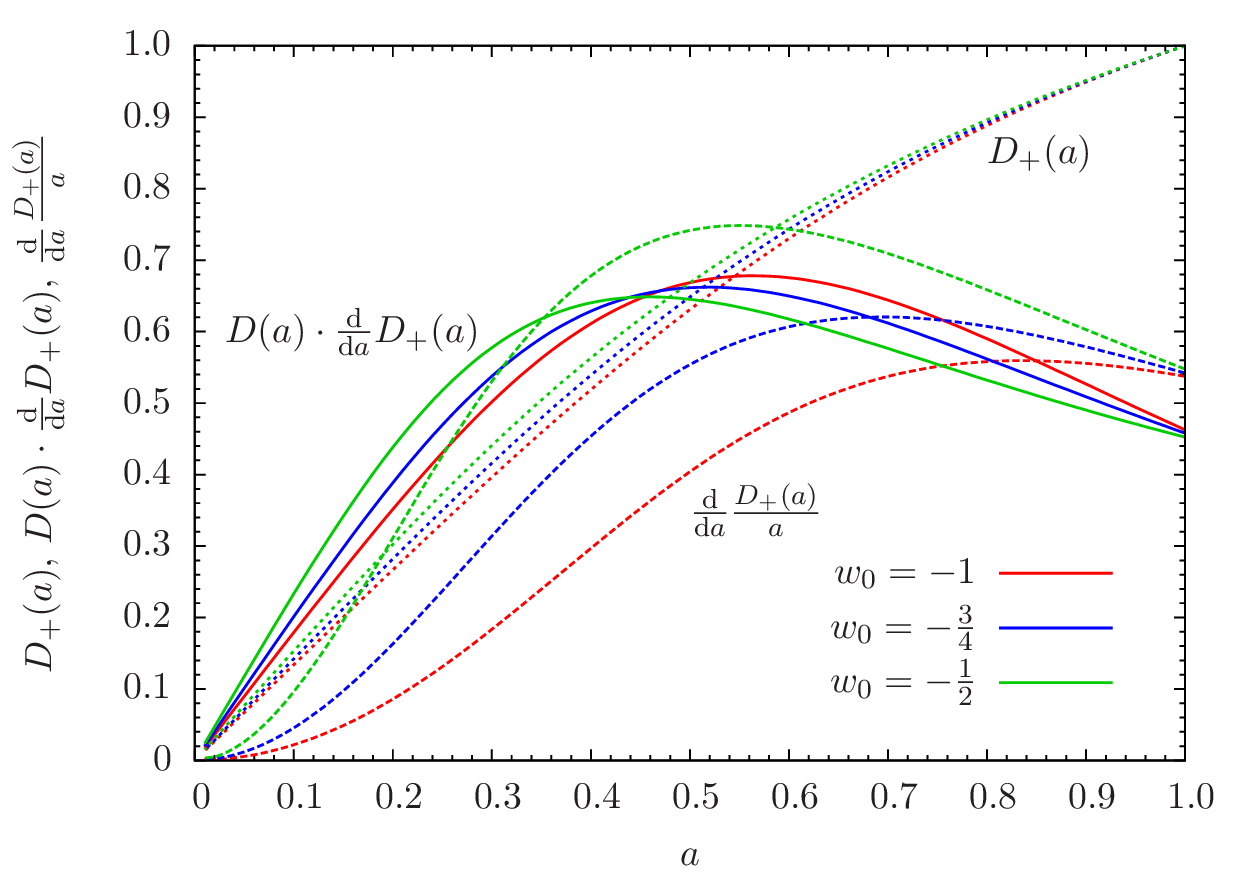}}
 \caption{Growth function of the linear (dashed line) and nonlinear (solid line) iSW-effect along with linear structure growth (dotted line), for varying equation of state parameters of the dark energy fluid.}
\end{figure}

So far, our derivation is essentially the same as the one presented by \citet{2002PhRvD..65h3518C}. \citet{2002PhRvD..65h3518C}, however, focused on the deeply nonlinear regime and proceeded by using approximations resulting from the halo model approach. In contrast to this, we are interested in an analytical description of the contributions to the nonlinear iSW-effect which arise from dark matter currents. Therefore, we restrict our analysis to the moderate nonlinear regime. 
Obviously, in this regime the simple relation between peculiar velocities and density contrast, given in equation~\eqref{eq_velocity_density_relation}, is no longer strictly valid. 
The CDM particles cease to exclusively follow the gradients of the potential.
However, in order to get an analytical estimate of the nonlinear contributions, it is still a useful approximation. Thus, we take nonlinear corrections to the continuity equation, manifest in the momentum density, into account, but we maintain the linear properties of the underlying fields, namely density contrast and peculiar velocities. Consequently, by virtue of equation~\eqref{eq_velocity_density_relation}, there is only one dynamical field, namely the density contrast, involved in our analysis. Equation~\eqref{eq_flux_in_Fourier_space} then may be written as
\begin{equation}
 \bmath j ( \kvec, \eta)= -\im D_+(\eta) \frac{\partial D_+(\eta)}{\partial\eta}
 	\int\frac{\dd^3k'}{(2\pi)^3}
	\frac{\bmath k - \bmath k'}{|\bmath k - \bmath k'|^2} \delta_0(\bmath k - \bmath k') \delta_0(\bmath k')
\end{equation}
omitting its linear part.
This expression for the current density has already been successfully used in the computation of the power spectrum of the Ostriker-Vishniac effect \citep{1986ApJ...306L..51O,1998PhRvD..58d3001J,2003PhRvD..67l3001C}.
Since in our framework all nonlinear contributions to the iSW-effect are solely sourced by the convolution part of the flux it is convenient to define its rescaled divergence
\begin{equation}
 \label{eq_def_Theta}
 \Theta(\kvec, \eta)\equiv - \frac{D_+(\eta)}{a} \frac{\partial D_+(\eta)}{\partial\eta}  
 \int\frac{\dd^3k'}{(2\pi)^3}
 \frac{\kvec\cdot(\bmath k - \bmath k')}{k^2|\bmath k - \bmath k'|^2} \delta_0(\bmath k - \bmath k') \delta_0(\bmath k')
\end{equation}
whereas the linear contributions are captured by
\begin{equation}
 \label{eq_def_tau_tilde}
 \tilde{\tau}(\kvec,\eta) \equiv \frac{\partial}{\partial \eta}\frac{D_+(\eta)}{a}\frac{\delta_0(\kvec)}{k^2}.
\end{equation}
With this the change of the gravitational potential takes the compact form
\begin{equation}
  \frac{\partial}{\partial \eta} \Phi (\kvec, \eta) =  -\frac{3}{2} H_0^2 \Omega_{\mathrm m} \left[\tilde{\tau}(\kvec,\eta) - \Theta(\kvec, \eta)\right].
\end{equation}
In $\Omega_{\mathrm m} / \chi^2_H$ we recognize the gravitational coupling constant mediating between the evolving matter fields and the resulting change in the gravitational potential.
Equations~\eqref{eq_def_Theta} and~\eqref{eq_def_tau_tilde} underline the different time evolution of the linear and nonlinear iSW-effect. For illustration we plot the relevant combinations of the growth function and its derivative for three different equation of state parameters of the dark energy component in Figure~\ref{fig_growth_function_and_derivatives}.

In order to characterize the statistics of the source field 
$\tau(\kvec,\eta)\equiv  \frac{\partial}{\partial \eta} \Phi (\kvec, \eta)$ 
we compute its power spectrum next
\begin{equation}
 \left\langle\tau(\kvec) \tau^*(\kvec') \right\rangle = (2\pi)^3 \delta_D(\kvec - \kvec') P_{\tau\tau}(k). 
\end{equation}
Because we assume the density contrast being a Gaussian random field its bispectrum vanishes and the power spectrum completely separates in its linear and first order nonlinear contributions
\begin{equation}
 P_{\tau\tau}(k) = P_{\tilde{\tau}\tilde{\tau}}(k) + P_{\Theta\Theta}(k).
\end{equation}
The linear contribution is given by $P_{\tilde{\tau}\tilde{\tau}}(k) = P_{\delta\delta}(k)/k^4$,
while for the evaluation of the second term we invoke Wick's theorem yielding
\begin{eqnarray}
\label{eq_convoultion_integral}
\nonumber
P_{\Theta\Theta}(k)&=&\int\frac{\dd^3k'}{(2\pi)^3}
 				\left[\frac{\kvec\cdot\left(\kvec-\kvec'\right)}{k^2|\kvec - \kvec'|^2} + \frac{\kvec\cdot\kvec'}{k^2k'^2}\right]
				\frac{\kvec\cdot(\kvec-\kvec')}{k^2|\kvec - \kvec'|^2}\\		
				&& \times P_{\delta\delta}\left(|\kvec -\kvec'|\right)P_{\delta\delta}(k').
\end{eqnarray}
At this point we would like to emphasize that our signal of the nonlinear iSW-effect completely originates from the trispectrum of the density field. In contrast to this, \citet{2002PhRvD..65h3518C} entirely neglects this term using arguments derived under the halo approach. Furthermore, he does not take terms involving cross-power spectra between the density and velocity fields into account. In our approach, however, these  terms do contribute since they are contained in the trispectrum. Another difference with respect to the work of \citet{2002PhRvD..65h3518C} results from its somewhat artificial discrimination between velocity-density correlations on the one hand and correlations of the density field and its (conformal) time derivative on the other hand. For the latter, \citet{2002PhRvD..65h3518C} borrows once more a result from the halo model assuming that for the fields involved the Cauchy-Schwarz inequality actually becomes an equality. In our approach, however, this far-reaching assumption can be avoided.

\begin{figure}
 \label{fig_weighting_functions}
 \centering
 \resizebox{\hsize}{!}{\includegraphics[]{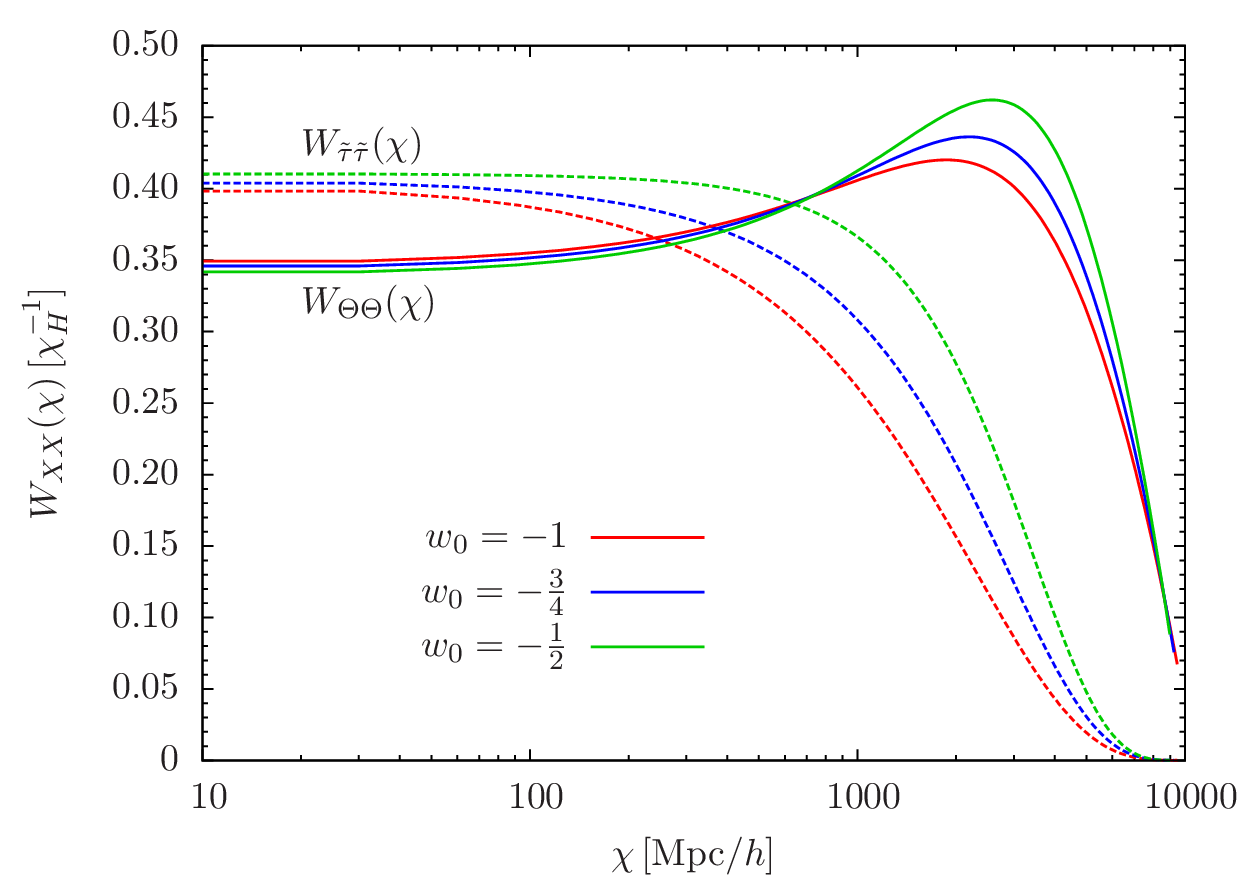}}
 \caption{Weighting function of the linear (dashed line) and nonlinear (solid line) iSW-effect for three different values of the equation of state parameter of the dark energy component.}
\end{figure}

The computation of the convolution integral in equation~\eqref{eq_convoultion_integral} is most readily carried out in a coordinate system where the wavevector $\kvec$ is aligned with the $x$-axis and by introducing spherical coordinates such that $\kvec'=\alpha k (\mu, \sin\vartheta \cos\phi, \sin\vartheta \sin\phi)$ with $\alpha\equiv k'/k$ and $\mu \equiv \cos\vartheta$. Expressing the CDM power spectrum in its dimensionless form, we find 
\begin{eqnarray}
 \nonumber
  \Delta^2_{\Theta\Theta}(k)
  &=& \frac{1}{2}\int_0^{\infty}\frac{\dd \alpha}{\alpha^2}\int_{-1}^1\dd\mu
  	\frac{(\alpha\mu-1) \left[\alpha\left(2\mu^2-1\right)-\mu\right]}{k^4\left(\alpha^2-2\alpha\mu+1\right)^{7/2}}\\
	&&\times \Delta^2\left(k\sqrt{\alpha^2-2\alpha\mu +1}\right)\Delta^2(\alpha k),
\end{eqnarray}
where we have defined $\Delta^2_{\Theta\Theta}(k)\equiv k^3P_{\Theta\Theta}(k)/2\pi^2$.
The corresponding angular power spectrum is obtained by Limber projection \citep{1953ApJ...117..134L,2001PhR...340..291B}
\begin{equation}
 \label{eq_Limber_projection}
 \frac{\ell(2\ell+1)}{4\pi}C^{XX}_\ell = \frac{\pi}{\ell}\int_0^{\chi_H}\chi\dd \chi\: W_{XX}^2(\chi)\Delta^2_{XX}\left(k=\ell/\chi\right) .
\end{equation}
In case of the linear iSW, the weight function is given by
\begin{equation}
 W_{\tilde{\tau}\tilde{\tau}}(\chi)=3\Omega_{\mathrm m}H_0^2 \frac{\partial}{\partial\eta} \frac{D_+(a)}{a},
\end{equation}
while for the first order nonlinear one we have
\begin{equation}
 \label{eq_weighting_function_theta_theta}
 W_{\Theta\Theta}(\chi) = 3\Omega_{\mathrm m}H_0^2 \frac{D_+(\eta)}{a} \frac{\partial}{\partial\eta}D_+(a).
\end{equation}
In Figure~\ref{fig_weighting_functions} we plot both weighting functions for varying equation of state parameter of the dark energy fluid. As expected, the weight function of the nonlinear iSW-effect peaks at comoving distances where the one of the linear iSW-effect has already started to decline rapidly. At high redshifts $\Omega_{\mathrm m}(a)$ approaches unity so that here growth function and scale factor coincide. Consequently, the linear iSW-effect vanishes identically (cf. equation~\ref{eq_def_tau_tilde}).

\begin{figure}
 \label{fig_power_spectra}
 \centering
 \resizebox{\hsize}{!}{\includegraphics[]{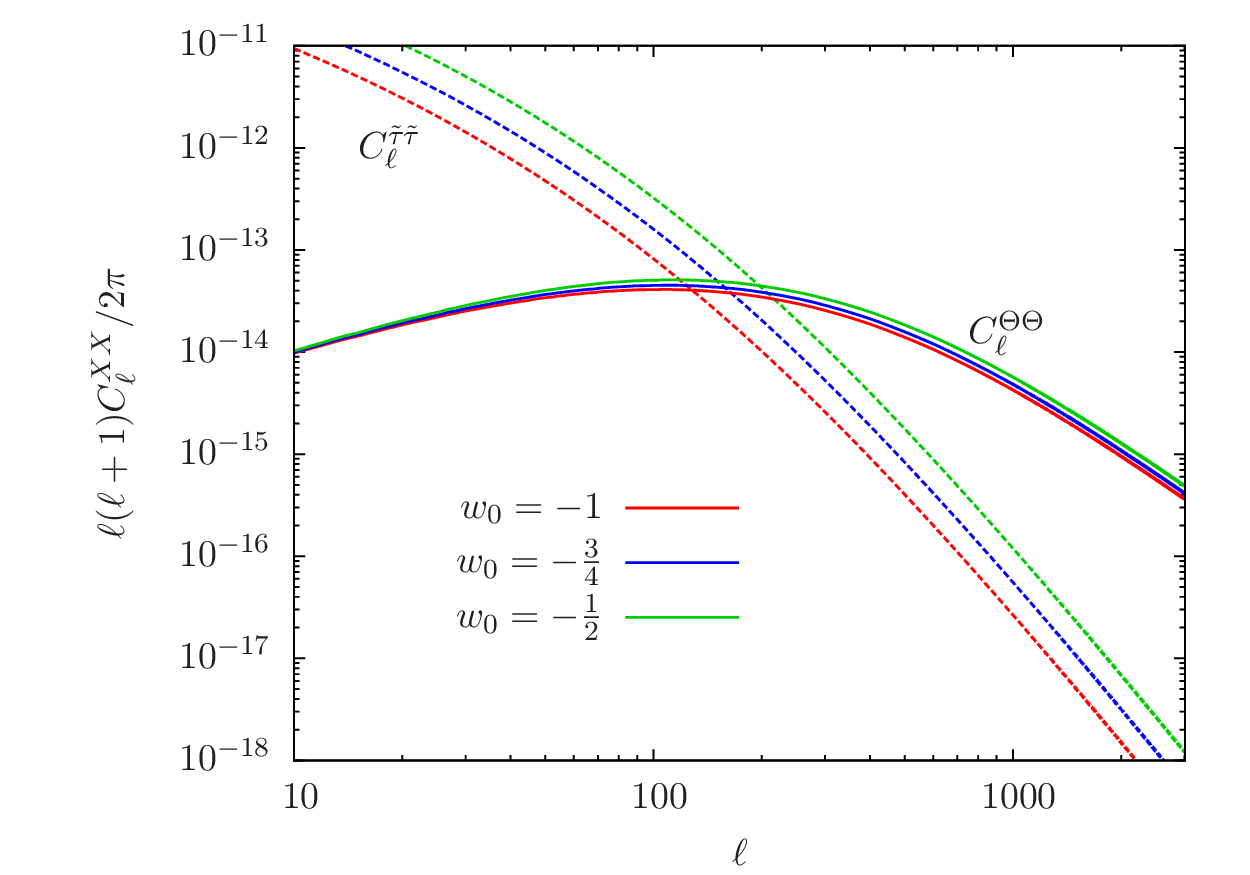}}
 \caption{Angular power spectra of the linear (dashed line) and nonlinear (solid line) iSW-effect for three different values of the equation of state parameter of the dark energy fluid.}
\end{figure}

The resulting power spectra are shown in Figure~\ref{fig_power_spectra}. 
Reflecting the additional fluctuation amplitude of the matter distribution on small spatial scales, the spectrum of the nonlinear iSW-effect is much flatter than the linear one. In case of the $\Lambda$CDM model  the nonlinear signal surpasses its linear counterpart at angular scales smaller than about one degree, i.e. $\ell\gtrsim100$. For time evolving dark energy models this crossing has shifted to smaller scales. However, being of second order in the density contrast, the amplitude of the nonlinear spectrum is much less sensitive to the underlying dark energy equation of state than the linear one.
Remarkably, the spectrum of the nonlinear iSW-effect presented here is very similar to the result of \citet{2002PhRvD..65h3518C} regarding its shape and amplitude, even though \citet{2002PhRvD..65h3518C} worked in the deeply nonlinear regime while our analytical estimate is dedicated to the translinear regime.

\begin{figure}
 \label{fig_smoothed_power_spectra}
 \centering
 \resizebox{\hsize}{!}{\includegraphics[]{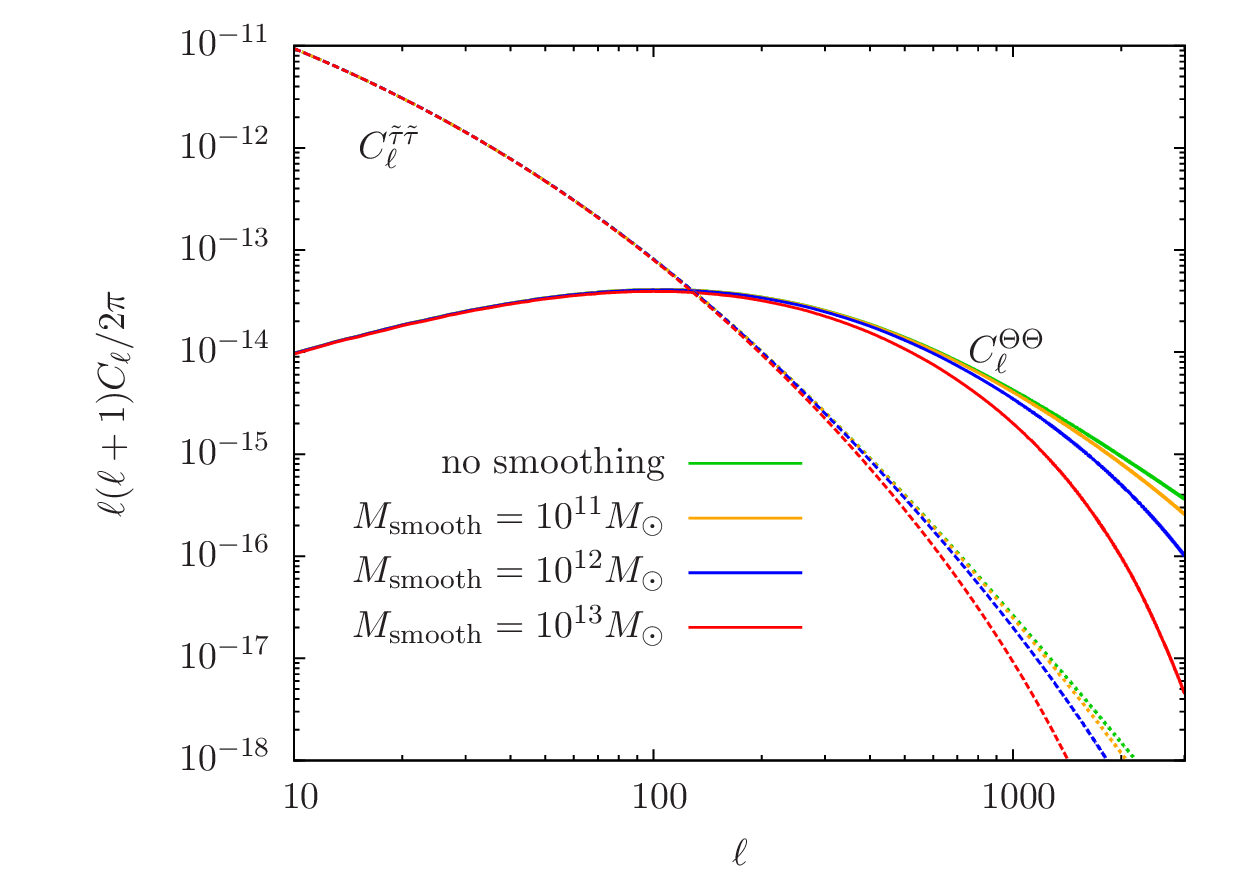}}
 \caption{Contributions of different mass scales to the angular power spectra of the linear (dashed line) and nonlinear (solid line) iSW-effect.}
\end{figure}

It is interesting to also investigate the mass dependence of the nonlinear iSW-effect. We content ourself to the $\Lambda$CDM model and smooth the power spectrum of the density contrast on different mass scales by introducing a Gaussian filter function
\begin{equation}
 P_{\delta\delta}(k) \longrightarrow P_{\delta\delta}(k)S^2_R(k) \qquad \mathrm{with} \qquad S_R(k) = \exp(-k^2R^2/2).
\end{equation}
The smoothing scale $R$ is set by the corresponding mass scale $M_{\mathrm{smooth}} = \frac{4\pi}{3}\rho_{\mathrm{crit}}\Omega_{\mathrm{m}}R^3$.
In Figure~\ref{fig_smoothed_power_spectra} we show the angular power spectra of the linear and nonlinear iSW-effect for three different masses, namely $10^{11},\, 10^{12}$ and $10^{13}$ solar masses. Excluding objects with masses larger than $10^{13}$ solar masses the power decreases considerably on small scales ($\ell \gtrsim 1000$), demonstrating that mostly cluster-size objects contribute to the signal.

\section{Contributions to the nonlinear integrated Sachs-Wolfe effect}
\label{sec_contributions}

A deeper analysis of equation~\eqref{eq_def_Theta} reveals some physical insights into the origin of the nonlinear iSW-effect. Mainly, one can identify two distinct sources. First, we rewrite equation~\eqref{eq_def_Theta} as the sum of two new fields
\begin{equation}
 \Theta(\kvec,\eta) = \Theta_{SG}(\kvec,\eta) + \Theta_{BG}(\kvec,\eta),
\end{equation}
with
\begin{equation}
 \Theta_{SG}(\kvec,\eta) \equiv -\frac{D_+(\eta)}{a} \frac{\partial D_+(\eta)}{\partial\eta}
 						\int \frac{\dd^3k'}{(2\pi)^3}
						\frac{\delta_0(\kvec - \kvec')}{|\kvec - \kvec'|^2} \delta_0(\kvec')
\end{equation}
and
\begin{equation}
 \label{eq_def_Theta_BG}
 \Theta_{BG}(\kvec,\eta) \equiv \frac{D_+(\eta)}{a} \frac{\partial D_+(\eta)}{\partial\eta}
 						\int \frac{\dd^3k'}{(2\pi)^3}
						\frac{\kvec \cdot \kvec'}{|\kvec - \kvec'|^2} \delta_0(\kvec - \kvec') \delta_0(\kvec').
\end{equation}
Concentrating on the first term, we proceed by making use of the Poisson equation and obtain
\begin{eqnarray}
 \nonumber
 \Theta_{SG}(\kvec,\eta) &=& 
 				\frac{H_0^{-2} \Omega_{\mathrm{m}}^{-1}}{3} 
				\frac{1}{a}\frac{\partial D^2_+(\eta)}{\partial\eta}
				\int\frac{\dd^3 k'}{(2\pi)^3} \Phi_0(\kvec - \kvec') \delta_0(\kvec')\\
 				\nonumber
				&=& 	
				\frac{H_0^{-2} \Omega_{\mathrm{m}}^{-1}}{3} 
				\left[ \frac{\partial}{\partial\eta} + a H(a)\right]\\
				&&
				\times
				\int\frac{\dd^3 k'}{(2\pi)^3} \Phi(\kvec - \kvec',\eta ) \delta(\kvec',\eta)
\end{eqnarray}
likewise in position space
\begin{equation}
\label{eq_theta_BG_rewritten}
 \Theta_{SG}(\x,\eta) = \frac{H_0^{-2} \Omega_{\mathrm{m}}^{-1}}{3}
				\left[ \frac{\partial}{\partial\eta} + a H(a)\right]
 				\Phi(\x,\eta) \delta(\x,\eta).
\end{equation}
Keeping in mind that we have already absorbed the inverse of the prefactor in the definition of the weighting function (cf. equation~\ref{eq_weighting_function_theta_theta}), the product $\Phi(\x)\delta(\x)$ can be interpreted as the gravitational self-energy density associated with the density contrast and its own gravitational potential. Consequently, the field $\Theta_{SG}$ may be interpreted as the (conformal) change of the gravitational self-energy density of the LSS together with the variation arising from the homogeneous expansion of the background. 

The interpretation of the second term is more involved. First, we rewrite equation~\eqref{eq_def_Theta_BG} in a more suggestive form
\begin{equation}
 \Theta_{BG}(\kvec,\eta) = \int\frac{\dd^3k'}{(2\pi)^3}
 \kvec'\frac{D_+(\eta)}{a}\frac{\delta_0(\kvec -\kvec')}{|\kvec-\kvec'|^2}
 \cdot\frac{\partial D_+(\eta)}{\partial\eta} \frac{\kvec}{k^2} \delta_0(\kvec').
\end{equation}
Using the Poisson equation, we recognize that the first term has the form of a gradient of the potential. The second term, however, resembles the velocity field (cf. equation~\ref{eq_velocity_density_relation}). Thus, we are tempted to write 
\begin{equation}
 \label{eq_theta_BG_real_space}
 \Theta_{BG}(\x,\eta) \simeq -2\nabla \Phi(\x,\eta)\cdot \bupsilon(\x,\eta)
\end{equation}
ignoring that equation~\eqref{eq_theta_BG_rewritten} does not fully obey the structure of a convolution. Note that we have also restored the prefactor contained in the weight function. Although being rather symbolic equation~\eqref{eq_theta_BG_real_space} has strong illustrative power. In the first term we recognize the lensing deflection angle $\balpha = -2\nabla_\bot \Phi$ \citep{2001PhR...340..291B}. Since the CMB temperature fluctuations are measured along the line of sight, only the three-dimensional gradient perpendicular to the light ray is considered. Accordingly, the scalar product just projects out the velocity components in the plane of the sky. Hence, we recover the dipole-like temperature anisotropy pattern which is usually associated with moving gravitational lenses \citep{1983Natur.302..315B, 1986Natur.324..349G}. The close relation between this effect and the nonlinear iSW-effect has already been pointed out by \citet{2002PhRvD..65h3518C}. However, neglecting the divergence part of the velocity field \citet{2002PhRvD..65h3518C} does not account for the contributions of the gravitational self-energy density we discussed before.

\begin{figure}
 \label{fig_three_d_spectra_different_contributions}
 \centering
 \resizebox{\hsize}{!}{\includegraphics[]{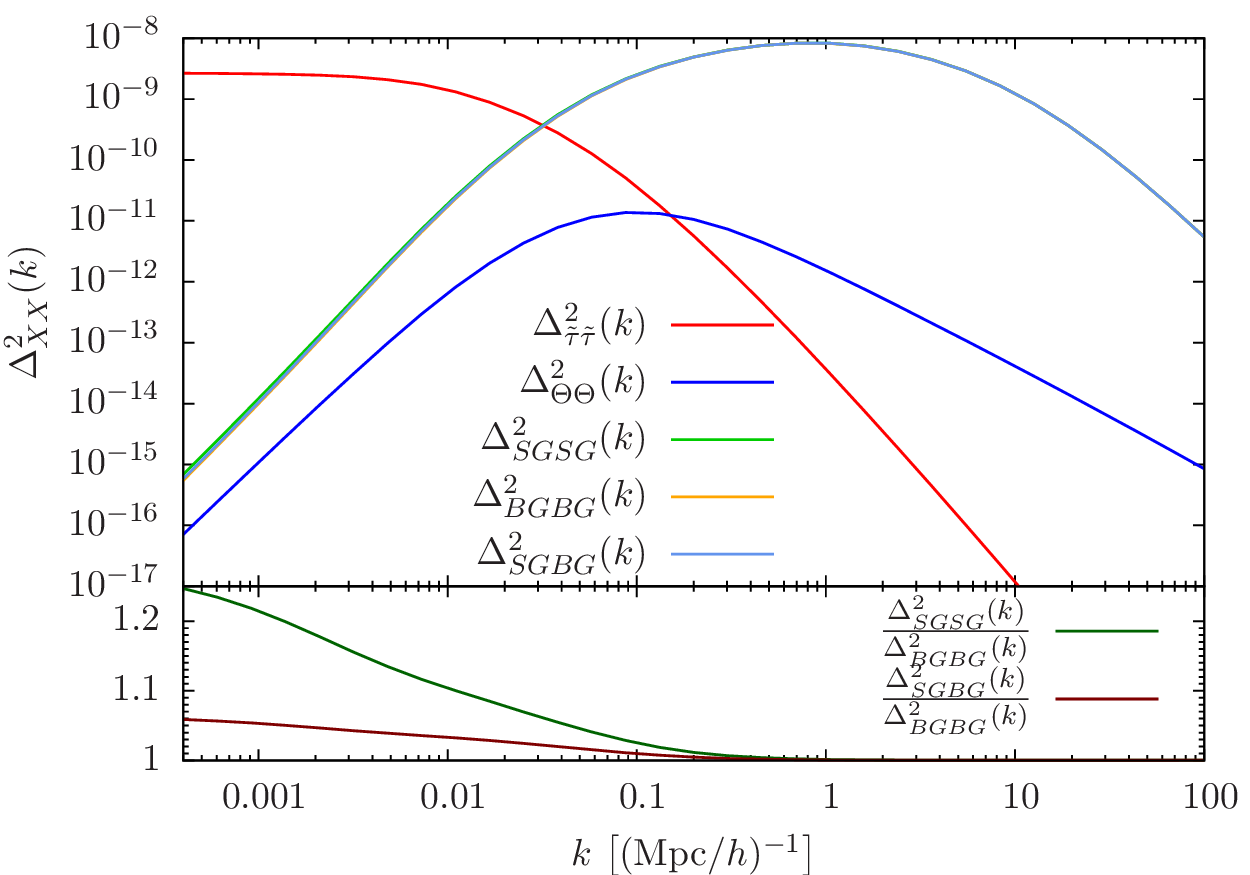}}
 \caption{Three dimensional power spectra of the gravitational self-energy density and Birkinshaw-Gull term constituting the nonlinear iSW-effect 
 as well as their cross spectrum (upper panel). Relative strength of the signal in comparison to the Birkinshaw-Gull term (lower panel).}
\end{figure}

\begin{figure}
 \label{fig_angular_spectra_different_contributions}
 \centering
 \resizebox{\hsize}{!}{\includegraphics[]{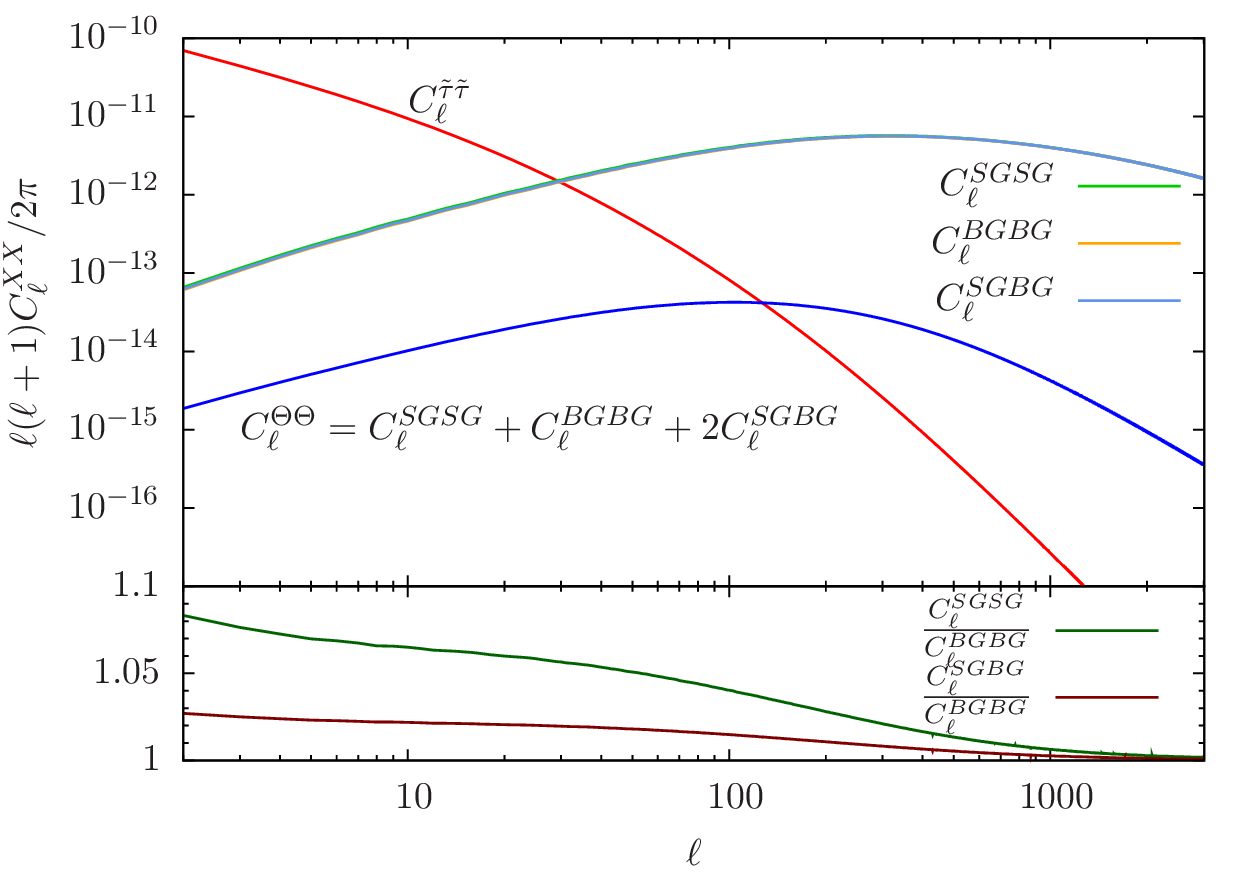}}
 \caption{Angular power spectra of the gravitational self-energy density and Birkinshaw-Gull term constituting the nonlinear iSW-effect as well as their cross spectrum (upper panel). Relative strength of the signal in comparison to the Birkinshaw-Gull term (lower panel).}
\end{figure}

For the computation of the power spectra of the two different contributions to the nonlinear iSW-effect we proceed in complete analogy to Section~\ref{sec_non_linear_isw}. In terms of the spectra of the two new fields $\Theta_{SG}$ and $\Theta_{BG}$ the power spectrum reads
\begin{equation}
 \label{eq_three_d_spectrum_contributions_combined}
 \Delta^2_{\Theta\Theta} (k) = \Delta^2_{SGSG}(k) + \Delta^2_{BGBG}(k) + 2\Delta^2_{SGBG}(k).
\end{equation}
The three different spectra involved can be jointly expressed as
\begin{eqnarray}
\nonumber
\Delta^2_{XX} (k) &=& \frac{1}{2} \int_0^{\infty} \dd \alpha \int_{-1}^{1} \dd\mu \;
				f_{XX} (k,\alpha,\mu) \Delta^2(\alpha k)\\
			&&	
			\times \frac{\Delta^2\left(k\sqrt{\alpha^2-2\alpha\mu+1}\right)}{\alpha (\alpha^2-2\alpha\mu +1)^{3/2}}
\end{eqnarray}
with
\begin{equation}
 f_{SGSG} (k,\alpha,\mu) = \frac{2\alpha^2 - 2\alpha\mu + 1}{\alpha^2k^4 ( \alpha^2-2\alpha\mu +1)^2 },
\end{equation}
\begin{equation}
 f_{BGBG} (k,\alpha,\mu) = \frac{\mu\left(
 							2\alpha^2 + 2\alpha^2\mu^2 - \alpha^3\mu -3\alpha\mu +1
 							\right)}{\alpha k^4 ( \alpha^2-2\alpha\mu +1)^2 }
\end{equation}
and the cross-spectrum
\begin{equation}
 f_{SGBG} (k,\alpha,\mu) = -\frac{\alpha \left(
 								2\alpha\mu^2 +\alpha - 3\mu
 								\right) + 1 }{\alpha^2k^4 ( \alpha^2-2\alpha\mu +1)^2 }.
\end{equation}
In Figure~\ref{fig_three_d_spectra_different_contributions} we plot the three-dimensional power spectra of the gravitational self-energy density and the Birkinshaw-Gull term along with the cross spectrum of both effects. In addition, we show the resulting power spectrum of the nonlinear iSW-effect and, for comparison, also the corresponding spectrum of the linear iSW-effect in the same figure.
From scales $k\gtrsim 0.1\, (\mathrm{Mpc}/h)^{-1}$ on the spectra of the $\Theta_{SG}$ and $\Theta_{BG}$ field, as well as their cross correlation, exceed the linear signal by several orders of magnitude. However, all three spectra are almost identical, as is demonstrated in the lower panel of Figure~\ref{fig_three_d_spectra_different_contributions}. Only on largest scales the relative difference amounts to 10-20\%. Therefore, their combination according to equation~\eqref{eq_three_d_spectrum_contributions_combined} results in the much lower signal of the nonlinear iSW-effect, which starts to surpass the spectrum of the linear effect on scales $k \sim 0.2\, (\mathrm{Mpc}/h)^{-1}$. Furthermore, we recognize that the three spectra peak at $k \sim 1\, (\mathrm{Mpc}/h)^{-1}$, whereas the maximum of their combination 
is shifted to larger scales ($k\sim0.1\, (\mathrm{Mpc}/h)^{-1}$). 
Limber projection (cf. equation~\ref{eq_Limber_projection}) of the three different spectra results in the corresponding angular power spectra shown at the top of Figure~\ref{fig_angular_spectra_different_contributions}.
Since the projection distributes the power of one mode over a wide multipole range the differences in the angular spectra are even smaller than in case of the three-dimensional power spectra (see lower panel of Figure~\ref{fig_angular_spectra_different_contributions}). From multipoles $\ell\gtrsim 1000$ all three spectra are essentially indistinguishable.

\section{Physical interpretation}
\label{sec_physical_interpretation}

To further motivate our interpretation of the contributions to the nonlinear iSW-effect we now present some analogies to the theory of gravitomagnetic potentials and classical field theory.

We start with the temperature anisotropy induced by a small lens moving with small (transverse) three-velocity $\bupsilon$. At lowest order we have
\citep{1983Natur.302..315B,1986Natur.324..349G} 
\begin{equation}
\frac{\Delta T_{BG}}{T_{\mathrm{CMB}}}= -2 \int\dd \chi\, \bupsilon\cdot \nabla \Phi,
\end{equation}
where the integral is evaluated along the photon path. Albeit treating the involved fields in linear theory, the Birkinshaw-Gull effect is obviously nonlinear in the sense that the perturbed quantities enter as a product.
In linear theory the velocity field can be obtained as the gradient of the potential (cf. equation~\ref{eq_velocity_density_relation}). Consequently, the Birkinshaw-Gull effect is of order $\mathcal O([\nabla \Phi]^2)$. 
In perturbation theory we can naturally construct one more term of this order, namely $\Phi\Delta \Phi$. By virtue of the Poisson equation this is exactly the gravitational self-energy density we have found in 
equation~\eqref{eq_theta_BG_rewritten}.

For simplicity we drop any prefactors in the following. Then, both effects can be obtained
from the auxiliary vector potential $\bmath A \equiv \Phi \nabla \Phi$ by taking the divergence
\begin{equation}
 \frac{\Delta T_{\mathrm{nl. \ iSW}}}{T_\mathrm{CMB}} = -2\int\dd\chi\, \mathrm{div}\, \bmath A = -2\int\dd\chi\, \left(\Phi \Delta \Phi + (\nabla \Phi)^2\right).
\end{equation}
Rewritting the potential $\bmath A$ as
\begin{equation}
 \bmath A(\x) = \int_{\mathcal V}\dd^3x' \frac{\delta(\x')\nabla \Phi(\x)}{|\x-\x'|}
\end{equation}
one immediately recognizes its similarity to the gravitomagnetic potential
\begin{equation}
 \bmath A_{\mathrm{GM}}(\x) \equiv \int_{\mathcal V} \dd^3 x' \frac{\bmath j (\x')}{|\x - \x'|} 
 \simeq\int_{\mathcal V}\dd^3x' \frac{\delta(\x') \nabla' \Phi(\x')}{|\x-\x'|}.
\end{equation}
The vector potentials can be identified when the change of the gravitational potential may be considered as being constant over the volume of integration $\mathcal V$. As long as both vector potentials coincide, the nonlinear iSW-effect is completely described by the Birkinshaw-Gull effect, as well as the gravitational self-energy term (at the corresponding order in $\Phi$) since \citet{2006MNRAS.369..425S} showed that $\Delta T_{\mathrm{nl. \ iSW}} / T_\mathrm{CMB} = 2\int\dd\chi\, \mathrm{div}\, \bmath A_{\mathrm{GM}}$. There is an important difference with respect to the theory of electromagnetism. In general, the vector potential describing the magnetic field is not fully determined by the scalar potential of the electric field. The scalar potential only enters implicitly via the charge density. At our level of approximation, however, not only the mass density $(\delta\sim \Delta\Phi)$ enters the gravitomagnetic potential but the velocity field is derived from the scalar potential, too $(\bupsilon \sim \nabla \Phi)$. Thus, given the scalar gravitational potential, the gravitomagnetic vector potential is already completely fixed.

It is interesting to note that the contribution of the self-energy density is independent of the assumption $\nabla \Phi \simeq \mathrm{const.}$ for $\x \in \mathcal V$. This can be seen as
\begin{eqnarray}
 \nonumber
 	\nabla \cdot \int_{\mathcal V}\dd^3x' \frac{\delta(\x') \nabla' \Phi(\x')}{|\x-\x'|}
 &=&
	\nabla \cdot \int_{\mathcal V}\dd^3x' 
	\biggl(
 		\Phi(\x') \nabla \frac{1}{|\x - \x'|}
	\biggr.
 \\
 &&
 \nonumber
	\biggl. 
		+\delta(\x') \nabla' \frac{\Phi(\x')}{|\x-\x'|}
	\biggr)\\
 &=&
 \nonumber
 	-4\pi\delta (\x) \, \Phi (\x)
 \\
 &&
	+ \nabla \cdot \int_{\mathcal V}\dd^3x' \delta(\x') \nabla' \frac{\Phi(\x')}{|\x-\x'|},
\end{eqnarray}
where we have first integrated by parts and subsequently interchanged the order of differentiation and integration.

The strong similarity to the potential $\bmath A$ also indicates why the statistics of the Birkinshaw-Gull and the self-energy density term are almost identical (cf. Figures~\ref{fig_three_d_spectra_different_contributions} and~\ref{fig_angular_spectra_different_contributions}). Integrated over a large volume one expects the divergence of $\bmath A$ to vanish due to Stokes' theorem. Accordingly, both terms, considered over a large volume, must cancel. Thus, one expects that they reveal the same statistical properties, whereas the small deviations point to the actual differences in the two vector potentials $\bmath A$ and $\bmath A_{\mathrm{GM}}$.

Finally, it is worth mentioning an analogy 
with the theory of massless scalar fields. Interpreting the potential $\Phi$ as scalar field interacting with an additional scalar field $\delta$
one readily computes its energy momentum tensor \citep[cf.][]{2000cils.book.....L}
\begin{equation}
 T_{\mu\nu} =  \partial_\mu \Phi\, \partial_\nu \Phi - \left( \frac{1}{2} \partial_\lambda \Phi \, \partial^\lambda \Phi + \delta\,\Phi\right) \eta_{\mu\nu}
\end{equation}
where $\eta_{\mu\nu}=\mathrm{diag}(-1,1,1,1)$ and we apply Einstein's summation convention.
Then the so-called \emph{mechanical pressure} $P$ of the field $\Phi$ is given by one third of the trace of the spatial part of the energy momentum tensor, i.e.
\begin{equation}
P \equiv \frac{1}{3}T^i_{\phantom{i}i} = \frac{1}{2}\dot\Phi^2-\frac{1}{6} (\nabla\Phi)^2 - \delta\,\Phi.
\end{equation}
We now see that, at our level of approximation, the nonlinear iSW-effect measures to some extent the mechanical pressure of the gravitational potential without the contribution arising from
its variation in time. That the term including time derivatives does not contribute is somewhat expected since in our derivation of the nonlinear iSW-effect we used the continuity equation to express the time variation of the gravitational potential by its variation in space.
The analogy with the mechanical pressure once more shows that the nonlinear iSW-effect is naturally constituted by both the Birkinshaw-Gull effect, as well as the gravitational self-energy density. The halo model approach of \citet{2002PhRvD..65h3518C}, however, cannot resolve this close relationship between these two contributions.

At the end of this section we shall comment on the apparent weakness of our approach. Major concerns may arise from the fact that we analyse a nonlinear effect starting from linear theory. We exploit the commonly used ansatz in perturbation theory that a product of first-order fields yields a second-order perturbative, i.e. nonlinear quantity. This approach, however, reveals at the same time one of the main problems of perturbation theory: it is not clear how to generalize this ansatz to higher order. On the other hand, this disadvantage is compensated for by the extreme simplification provided by this ansatz. It allows for the reduction to one single dynamical field, the density contrast or likewise the gravitational potential. One can then easily identify the terms which contribute at the same order in the perturbative field. Furthermore, the statistical properties may be derived most conveniently by tracing back higher order correlators to the matter power spectrum (cf. Section~\ref{sec_contributions}). Thus, our approach facilitates a fully analytical treatment, which in turn offers some direct physical interpretations of the origin of the nonlinear iSW-effect. These interpretations are widely inspired by analogies and always prefer the illustrative power of physical reasoning.

\section{Summary}
\label{sec_summary}

In this paper we revisited the theory of the nonlinear iSW-effect. We used the continuity equation to express the time evolution of the gravitational potential in terms of the divergence of the dark matter flux density. The momentum density, being the product of density and velocity field, can then be interpreted as source field of the nonlinear iSW-effect.
\begin{enumerate}
	\item Our ansatz treats both dynamical fields, density contrast and velocity, in linear theory. Consequently, the velocity field is completely 
		determined by the density contrast, leaving the latter as the only dynamical quantity involved in our analysis. This simplification allows 
		for a fully analytical treatment in contrast to the previous work of \citet{2002PhRvD..65h3518C} which substantially relies on results from the 
		halo model.
	\item Exploiting Wick's theorem we derived the angular power spectrum of the nonlinear iSW-effect. We confirmed the shape and amplitude found 
		by other authors using different approaches \citep{1996ApJ...460..549S,2002PhRvD..65h3518C}. Especially, we verified that the nonlinear 
		signal surpasses that of the linear iSW-effect at multipoles $\ell \gtrsim 100$.
	\item Our analytical ansatz allowed to reveal two contributions to the nonlinear iSW-effect: the Birkinshaw-Gull effect \citep[as already pointed out 
		by][]{2002PhRvD..65h3518C} and the conformal change of the gravitational self energy density of the cosmic large scale structure.
	\item Computing the three-dimensional power spectra and the corresponding angular power spectra of the individual contributions, we found that 
		they are almost identical, except on large scales, and their individual detection is certainly impossible for ongoing CMB experiments.
	\item We showed by simple arguments from perturbation theory that it is natural that the Birkinshw-Gull term is accompanied by the gravitational 
		self-energy density term. Furthermore, we pointed out several analogies with the theory of gravitomagnetic potentials and scalar fields to 
		highlight the physical meaning of the different contributions to the nonlinear iSW-effect.
\end{enumerate}
Despite its confinement to linear theory our ansatz allows for a deeper understanding of the physical processes underneath the nonlinear iSW-effect, 
especially regarding its relation to other phenomena like the Birkinshaw-Gull effect.

\section*{Acknowledgements}

PhMM acknowledges funding from the Graduate Academy Heidelberg and support from the International Max Planck
Research School for Astronomy and Cosmic Physics in Heidelberg as well as from the Heidelberg Graduate School of Fundamental Physics.
BMS's work is supported by the German Research Foundation (DFG) within the framework of the excellence initiative through the Heidelberg Graduate School of Fundamental Physics.

\bibliography{bibtex/aamnem,bibtex/references}
\bibliographystyle{mn2e}


\bsp

\label{lastpage}

\end{document}